# Antiferromagnetic Stabilization in $Ti_8O_{12}$

Xiaohu Yu[1*], Artem R. Oganov[2,1,3,4], Guangrui Qian[3], Ivan A. Popov[5] and Alexander I. Boldyrev[5]

[1]*Department of Problems of Physics and Energetics, Moscow Institute of Physics and Technology, 9 Institutskiy Land, Dolgoprudny city, Moscow Region 141700, Russia*

[2]*Skolkovo Institute of Science and Technology, Skolkovo Innovation Center, Bldg. 5, Moscow 143026, Russia.*

[3]*Department of Physics and Astronomy, Stony Brook University, Stony Brook, New York 11794, USA*

[4]*School of Materials Science, Northwestern Polytechnical University, Xi'an 720072, China*

[5]*Department of Chemistry and Biochemistry, Utah State University, Logan, Utah 84322, United States*

*Abstract*

*Using the evolutionary algorithm USPEX and DFT+U calculations, we predicted a high-symmetry geometric structure of bare $Ti_8O_{12}$ cluster composed of 8 Ti atoms forming a cube, which O atoms are at midpoints of all of its edges, in excellent agreement with experimental results. Using Natural Bond Orbital analysis, Adaptive Natural Density Partitioning algorithm, electron localization function and partial charge plots, we find the origin of the particular stability of bare $Ti_8O_{12}$ cluster: unique chemical bonding where eight electrons of Ti atoms interacting with each other in antiferromagnetic fashion to lower the total energy of the system. The bare $Ti_8O_{12}$ is thus an unusual molecule stabilized by d-orbital antiferromagnetic coupling.*

*\* E-mail: yuxiaohu950203@126.com.*

Photoactivity of $TiO_2$[1] is one of the most attractive properties for technologies making use of titanium oxide, which include photoelectrochemical splitting of water to hydrogen[2], oxidation photocatalysis of organic molecules[3], photovoltaic devices[4] and super hydrophilicity[5]. Titanium oxide clusters and their derivatives have also attracted great attention owing to their remarkable properties[6]. X-ray diffraction studies show that titanium oxychloride $[Ti_8O_{12}(H_2O)_{24}]Cl_8 \cdot HCl \cdot 7H_2O$ has a cubic titanium octamer linked to chloride ions and water molecules by a complex network of hydrogen bonds[7]. Titanium oxychloride is useful as a semiconductor element of a photovoltaic cell or as a photocatalyst in air or water purification treatments[8]. In addition, single crystals of titanium oxychloride can be synthesized by gentle hydrolysis of a commercial $TiOCl_2$ solution, and hydrolysis of the titanium oxychloride can be used for synthesis of various $TiO_2$ materials[9]. Theoretically, $Ti_8O_{12}H_8$ cluster can be used as an ideal model to study transformation of methanol to formaldehyde[10]. The natural question is: what is the origin of stability of the bare $Ti_8O_{12}$ cluster and the $Ti_8O_{12}$ complex? Usually, the geometric structure and particularly favorable chemical bonding patterns are associated with the stability of such clusters. There are many reports on unusual chemical bonding modes in stable transition metal oxide clusters[11], e.g. $(Mo_3O_9)^{2-}$ and $(W_3O_9)^{2-}$ with d-orbital σ aromaticity[12]. Unique π- and δ-aromaticity in transition metal oxide cluster $(Ta_3O_3)^-$ is also known[13]. Very recently, d-orbital spherical σ-aromaticity in a stable $Ce_6O_8$ cluster was found by Yu et al.[14]. In this communication we report an interesting and unexpected chemical bonding mode in bare $Ti_8O_{12}$ cluster exhibiting d-orbital antiferromagnetic character, which is responsible for high stability of the bare $Ti_8O_{12}$ cluster.

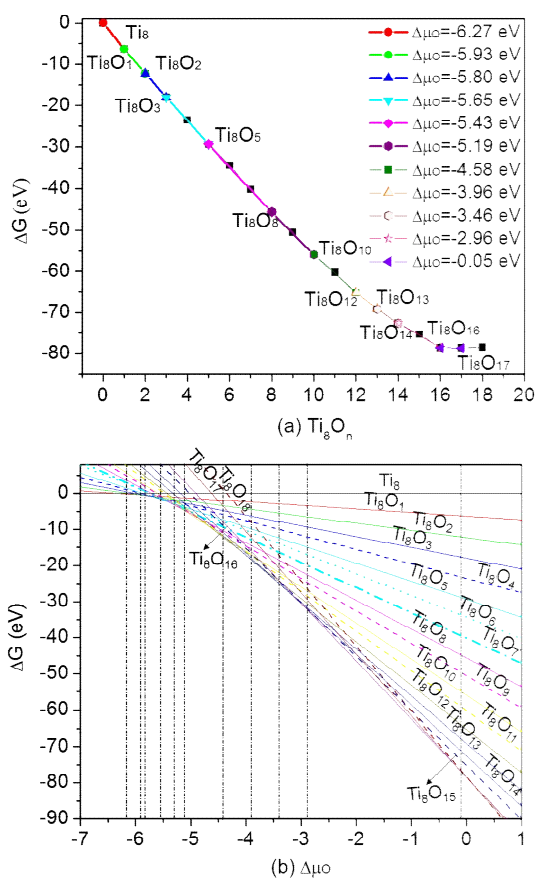

**Figure 1.** Phase diagram of $Ti_8O_n$ (n=1-18) clusters: stability of different clusters as a function of (a) number of oxygen atoms and (b) oxygen chemical potential.

Extensive searches for the global minima of all $Ti_8O_n$ (n=1-18) clusters were performed at the spin-polarized DFT+U level of theory. The predicted global energy minimum of the $Ti_8O_{12}$ cluster has titanium atoms in the corner of the $Ti_8$ cube and oxygen atoms at midpoints of its edges, which agrees well with experiment[7]. The Ti-Ti, O-O and Ti-O distances are 3.18, 3.22, 1.86 Å, respectively. The structures and relative total energies for the ground electronic state and low-lying isomers of $Ti_8O_{12}$ are shown in Supporting Information Figure S1.

First, we analyzed stability of $Ti_8O_n$ (n=0-18) clusters using *ab initio* thermodynamics, with the Gibbs free energy defined as[15]:

$\Delta G_f(T, p_{O2}) = F_{Ti8On}(T) - F_{Ti8}(T) - n\mu_O(T, p_{O2})$  (1)

where $F_{Ti8On}(T)$ and $F_{Ti8}(T)$ are the Helmholtz free energy of the $Ti_8O_n$ and the pristine $Ti_8$ cluster (at their ground state with respect to geometry and spin), respectively, and $\mu_O(T, p_{O2})$ is the chemical potential of oxygen. There are twelve stable clusters: $Ti_8$, $Ti_8O_1$, $Ti_8O_2$, $Ti_8O_3$, $Ti_8O_5$, $Ti_8O_8$, $Ti_8O_{10}$, $Ti_8O_{12}$, $Ti_8O_{13}$, $Ti_8O_{14}$, $Ti_8O_{16}$ and $Ti_8O_{17}$ as shown in Figure 1a. We can recast Figure 1a in another insightful form as shown in phase diagram (Figure 1b). In equilibrium with molecular $O_2$ gas, $\mu_O(T, p_{O2})$ is expressed as:

$\mu_O(T, p_{O2}) = \frac{1}{2}[E_{O2} + \mu_{O2}(T, p^0) + k_BT\ln(p/p^0)]$,  (2)

where $p^0$, $k_B$, and p are the standard atmospheric pressure, Boltzmann constant, and oxygen partial pressure, respectively. $E_{O2}$ was obtained from a spin-polarized calculation. $\mu_{O2}(T, p^0)$ term includes vibrational, rotational, and magnetic contributions for $O_2$ gas, and was taken from thermodynamic tables[16]. $k_BT\ln(p/p^0)$ is the contribution of temperature and $O_2$ partial pressure to the oxygen chemical potential.

The phase diagram as a function of temperature and partial oxygen is shown in Figure 2 — it is clear that $Ti_8O_{12}$ has a wide stability field. Spin density distribution for stable $Ti_8O_n$ (n=10, 12, 13, 14, 16) clusters is also shown in Figure 2. The ground state of $Ti_8O_{16}$ cluster is singlet, $Ti_8O_{14}$ is quintet, and both $Ti_8O_{12}$ and $Ti_8O_{13}$ clusters are both singlets in antiferromagnetic state.

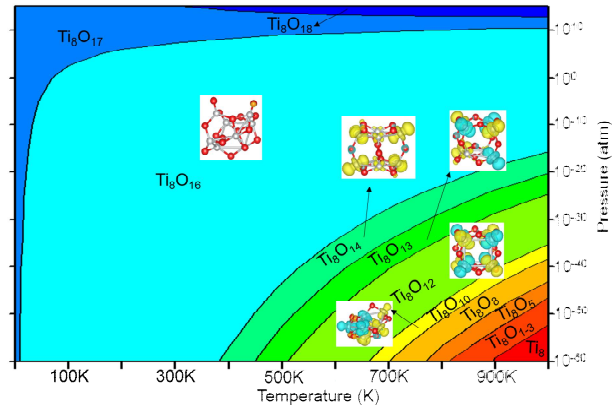

**Figure 2.** Phase diagram of $Ti_8O_n$ in oxygen atmosphere. The spin distribution (spin up: yellow; spin down: light blue) of the corresponding clusters is shown in the phase diagram.

Then, the second energy difference was used to determine stable clusters of $Ti_8O_n$, it is defined as:

$\Delta^2E = 2E(Ti_8O_n) - E(Ti_8O_{n-1}) - E(Ti_8O_{n+1})$,  (3)

Where $E(Ti_8O_n)$ is the energy of $Ti_8O_n$ cluster. The peaks of second energy difference (Figure 3) correspond to the magic clusters: $Ti_8O_{10}$, $Ti_8O_{12}$, $Ti_8O_{16}$.

The total oxygen binding energy is defined as:

$\Delta\Delta = \Delta E_n - \Delta E_{n-1}$,  (4)

and the averaged oxygen binding energy is $\Delta E = \Delta E_n/n$. From the stepwise oxygen binding energy (Figure 3), we see that the binding of up to sixteen oxygen atoms on the $Ti_8$ cluster is particularly favorable thermodynamically, and the averaged oxygen binding energy shows that binding of all eighteen oxygen atoms to the $Ti_8$ cluster is still thermodynamically feasible.

Stability of $Ti_8O_{12}$ and $Ti_8O_{16}$ is due to closed electronic shell, but $Ti_8O_{12}$ (on which we focus here) also has a highly symmetric atomic structure ($O_h$ symmetry), in contrast to all other $Ti_8O_n$ (n=1-18) clusters with lower point group symmetries (Figure S2). $Ti_8O_{12}$ complexes play an important role in crystallization of bulk $TiO_2$[9, 17]. There are many experimental reports on the synthesis of $Ti_8O_{12}$ complexes and a series of nano titanium oxides using $Ti_8O_{12}$ complex[17-18].

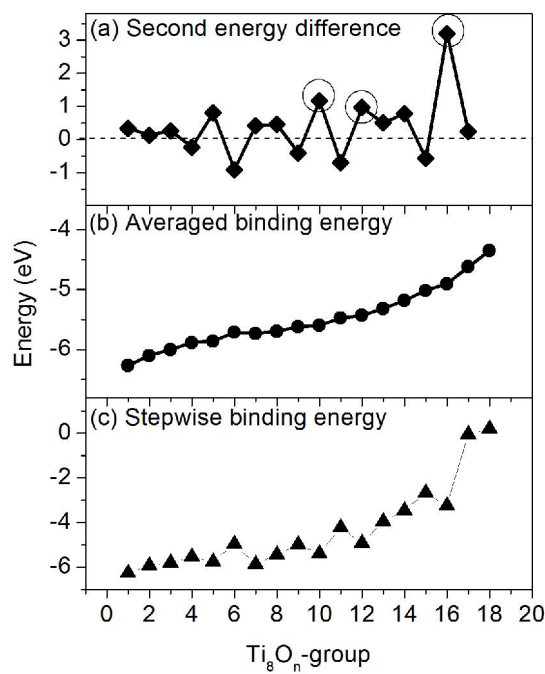

**Figure 3.** Second energy difference, averaged and stepwise binding energies for global minima $Ti_8O_n$ (n=1-18) clusters.

Counting valence electrons, 8 titanium atoms give a total of 8 × 4 = 32 electrons for bonding and 12 oxygen atoms can get in total 12 × 2 ($O^{2-}$) = 24 electrons. Hence, the formal charge on Ti is assumed to be +3. On balance, there are 32 − 24 = 8 electrons, responsible for the direct Ti-Ti bonding. Understanding where these remaining 8 electrons localize is very important for understanding the extra stability of $Ti_8O_{12}$ cluster.

Electron localization function (ELF)[19] proved to be a useful tool for studying chemical bonding in clusters and molecules. It is interesting to find that there is a localization of electron density on faces of $Ti_8O_{12}$ cluster, as shown in Figure S3. This is reminiscent of the newly predicted polyzinc compound containing $[Zn^I_8(HL)_4(L)_8]^{12-}$ (L=tetrazoledianion) cluster core[20], which is stable and also features a maximum of ELF in faces of $Zn_8$ cube, corresponding to an electron pair localized on each face of the cluster.

Figure 4 shows total and projected densities of states (DOS) of the $Ti_8O_{12}$ cluster. From DOS, we can clearly see that the Ti-4s, 3d, 3p hybridize and form bonds with oxygen atoms. Molecular orbitals in the range of -5 to -2 eV are made of strongly hybridized Ti-d, Ti-p and Ti-s, with O-p orbitals forming strongly bonding molecular orbitals. Most interesting bonding molecular orbitals (HOMO-1 state) with energies in the range from -1.2 to -0.8 eV are mainly composed by Ti-$d_{xy}$, $d_{yz}$, $d_{xz}$, with a small contribution from O-p orbitals. The other interesting molecular orbitals (HOMO state) with energies in the range from -0.3 to 0.3 eV are mainly composed by Ti-$d_{xy}$, $d_{yz}$, $d_{xz}$, and a little O-p, Ti-s, Ti-$d_{z^2}$. The energies in the range from 1.8 to 2.2 eV correspond to antibonding state, which is mainly composed of five d-orbitals of 8 Ti atoms and a small admixture of oxygen orbitals.

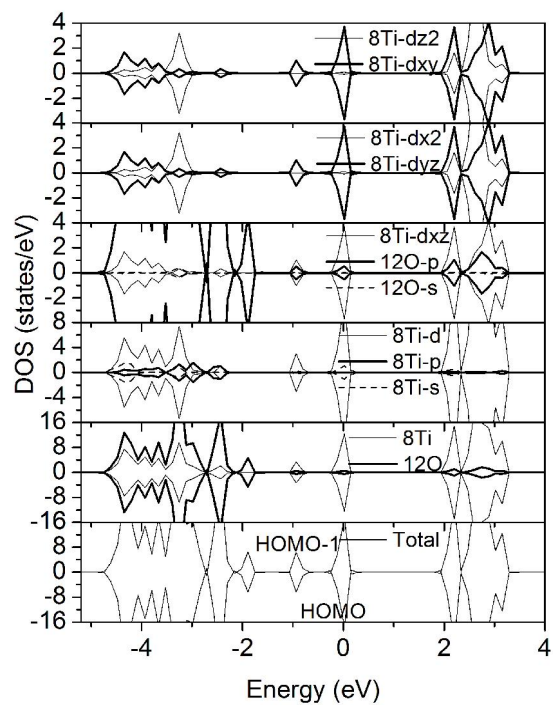

**Figure 4.** Total and projected densities of states (DOS) of the $Ti_8O_{12}$ cluster.

Figure 5 shows partial charge images of the $Ti_8O_{12}$ cluster. From both HOMO and HOMO-1 (Figure 5), one can see the electron density localization inside the $Ti_8$ fragment (also see Figure S4). To get more insight into the electronic structure of the $Ti_8O_{12}$ cluster, we used the Natural Bond Orbital (NBO) as well as its extension Adaptive Natural Density Partitioning Analysis (AdNDP) to further interpret the results described above. Previously, AdNDP results have been shown to be insensitive to the level of theory or basis set used[21]. While NBO allows determination of Lewis elements of localized bonding, such as 1c-2e bonds (lone pairs) and 2c-2e bonds (classical two-center two-electron bonds), AdNDP enables delocalized bonding (nc-2e bonds, n > 2) to be also found. In total, there are 52 valence electron pairs in the cubic $Ti_8O_{12}$ cluster. Similarly to the case of highly symmetric $Ce_6O_8$ cluster[14], where the localization of two d-electrons inside the $Ce_6$ fragment was found to be responsible for the stability of the cluster, one can assume that eight d-electrons coming from eight Ti atoms may also lead to d-orbital cubic σ aromaticity according to the 6n+2 or 2(n+1)$^2$ electron counting rule[21]. Indeed, NBO and AdNDP analyses at the DFT level of theory of a hypothetical non-magnetic $Ti_8O_{12}$ (where all its electrons are paired) shows a formation of the four 8c-2e σ bonds (Figure S5 and description), responsible for the d-AO based σ aromaticity, akin to the case of $Ce_6O_8$[14]. However, according to the global minimum search, cubic $Ti_8O_{12}$ is antiferromagnetic in its ground electronic state. Thus, ideally, there should be four unpaired α- and four unpaired β-electrons coming from 8 Ti atoms. Indeed, the complete active space self-consistent field (CASSCF) calculations showed a very strong multiconfigurational nature of the wavefunction (the first leading coefficient is ~0.3) and confirmed the antiferromagnetic character of the cluster (Table S1). Thus, the natural question is: what causes stabilization of the cluster? In fact, according to the DFT+U calculations, the antiferromagnetic state of the global minimum structure of $Ti_8O_{12}$ is 0.74 eV lower in energy than the ferromagnetic state, while CASSCF calculations show an even greater value of 1.60 eV. We further performed AdNDP analyses of the antiferromagnetic $Ti_8O_{12}$ using the CASSCF wavefunction to understand how the chemical bonding picture is changed compared to the non-magnetic case. Noteworthy, s-type and p-type lone pairs on oxygen atoms as well as all the 2c-2e Ti-O σ bonds were found to be exactly the same as for the non-magnetic cluster with just slight differences in the occupation number (ON) values. However, the remaining eight electrons were found to localize in accordance with the one-electron symbolic density matrix coefficients for the corresponding orbitals: $a_{1g}^{1.53}$ $t_{1u}^{3.42}$ $t_{2g}^{2.55}$ $a_{2u}^{0.49}$, where $a_{1g}$ and $t_{1u}$ orbitals are bonding, and $a_{2u}$ and $t_{2g}$ are their respective antibonding counterparts. In fact, the totally symmetric 8c-2e σ bond (Figure S5g) (originating from the bonding HOMO-1 ($a_{1g}$) orbital) was found to have lower ON value of 1.53 |e| instead of 1.89 |e|. Noteworthy, three 8c-2e σ bonds (Figure S5d-f) originating from the triply degenerate HOMO $t_{1u}$ orbital have appreciably lower ON values: 1.14 |e| instead of 1.87 |e|. Thus, one can conclude that there is indeed a certain stabilization coming from some d-AO overlap of 8 Ti electrons, but it cannot be called aromatic as in the case of the hypothetical non-magnetic structure with four bonding orbitals, each occupied by ~2 electrons. Indeed, the geometric factor plays an important role: 8 Ti atoms separated from each other by 12 O atoms are forced to adopt an antiferromagnetic configuration to maximize d-orbital interactions through superexchange, thus lowering the total energy of the system. It should also be pointed out that, primarily, system stabilization is achieved via the strong ionic interactions between $Ti^{3+}$ and $O^{2-}$. Although common in the solid state, antiferromagnetism in a small cluster is found here for the first time.

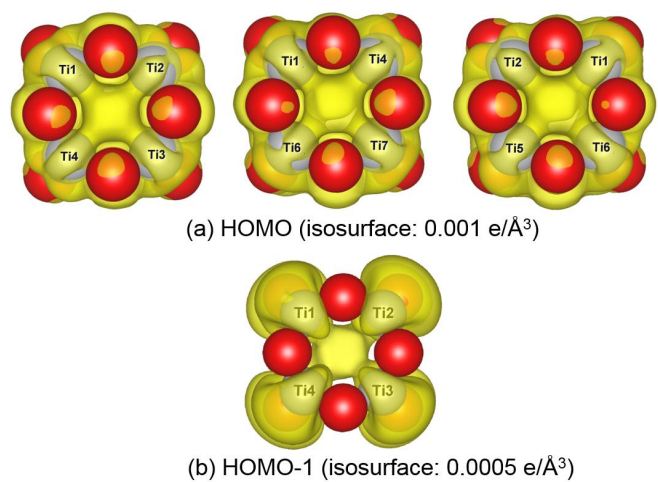

(a) HOMO (isosurface: 0.001 e/Å$^3$)

(b) HOMO-1 (isosurface: 0.0005 e/Å$^3$)

**Figure 5.** Partial charge images of $Ti_8O_{12}$ cluster (a) HOMO and (b) HOMO-1.

In conclusion, using evolutionary structure prediction code USPEX and DFT+U calculations we found a highly symmetric and stable transition metal oxide cluster $Ti_8O_{12}$. The phase diagram as a function of temperature and partial oxygen pressure indicates a wide stability field of the $Ti_8O_{12}$ cluster. From the analysis of its electron localization function and electronic structure of $Ti_8O_{12}$, we found that neutral $Ti_8O_{12}$ cluster is stabilized due to the strong ionic Ti-O bonding and d-orbital antiferromagnetic coupling. The peculiar chemical bonding is likely to make this cluster and its derivatives very interesting for catalysis. We hope that this work will inspire experimental realization of transition metal and lanthanide oxides clusters exhibiting d-orbital antiferromagnetic stabilization.

## Experimental Section

The *ab initio* evolutionary algorithm USPEX[22], which has been successfully applied to various materials[23], was used to look for the stable $Ti_8O_n$ (n = 1-18) clusters. More detailed description of these calculations can be found in the Supporting Information. To understand chemical bonding in the $Ti_8O_{12}$ cluster, we have utilized NBO [24] and AdNDP[25] analyses, both at PBE0/LANL2DZ and CASSCF(8,10)/LANL2DZ levels of theory using the geometry obtained from VASP calculations. AdNDP has been shown to be a very effective tool for the description of chemical bonding in various systems: from small clusters (0D)[26] to the chain-like species (1D)[27], various atomic-thin sheets (2D)[28], and hypervalent species[29].

## Acknowledgements

We acknowledge financial support from the Government of the Russian Federation (#14.A12.31.0003). This work was supported by the National Science Foundation (CHE-1361413 to A. I. B.) in the United States. Computer, storage, and other resources from the Division of Research Computing in the Office of Research and Graduate Studies at Utah State University are gratefully acknowledged.